\documentclass[aps,prb,twocolumn]{revtex4-1}
\usepackage{graphicx}
\usepackage{bm}
\usepackage{amsmath}
\usepackage{amssymb}
\usepackage{euscript}
\usepackage{verbatim}
\usepackage{fancyhdr}
\usepackage{wrapfig}
\usepackage{setspace}
\usepackage{xcolor}
\usepackage{amsfonts}
\usepackage{subfigure}

\newcommand{\be}{\begin{equation}}
\newcommand{\ee}{\end{equation}}
\newcommand{\bea}{\begin{eqnarray}}
\newcommand{\eea}{\end{eqnarray}}

\begin{document}

\begin{titlepage}

\title{Charge Fractionalization in a Kondo Device}
\author{L. Aviad Landau}
\affiliation{Raymond and Beverly Sackler School of Physics and Astronomy, Tel-Aviv University, IL-69978 Tel Aviv, Israel}
\author{Eyal Cornfeld}
\affiliation{Raymond and Beverly Sackler School of Physics and Astronomy, Tel-Aviv University, IL-69978 Tel Aviv, Israel}
\author{Eran Sela}
\affiliation{Raymond and Beverly Sackler School of Physics and Astronomy, Tel-Aviv University, IL-69978 Tel Aviv, Israel}

\begin{abstract}
We study nonequilibrium transport through a charge Kondo device realizing the two-channel Kondo critical point in a recent experiment by Iftikhar \textit{et al} \cite{iftikhar2016two}. By computing the current and shot noise at low voltages near the critical point, we obtain a universal Fano factor $e^*/e=1/2$. We identify elementary transport processes as weak scattering of emergent fermions carrying half-integer charge quantum numbers. This forms an experimental fingerprint for fractionalization in a non-Fermi liquid, which, compared to spin-Kondo devices, could be observed at elevated temperatures.

\end{abstract}


\maketitle

\draft

\vspace{2mm}

\end{titlepage}

\emph{Introduction and Results.--} Deconfinement and fractionalization are fascinating phenomena in which particles that are initially found as bound states become independent of each other. Such phenomena emerge in strongly interacting condensed matter systems, for example, in the form of spin-charge separation in Luttinger liqudis~\cite{auslaender2005spin}, possible emergence of a spinon Fermi sea in spin liquids~\cite{lee2008end}, and the appearance of magnetic monopoles in spin-ice~\cite{kadowaki2009observation}. Deconfinement is often associated with asymptotic freedom as occurring in gauge theories~\cite{gross1973ultraviolet} leading to the quark-gluon plasma in high energy physics.  Arguably, the simplest strongly interacting model that displays asymptotic freedom is the Kondo effect, describing magnetic impurities in metals, and revived in the 90's in the realm of quantum dots~\cite{kouwenhoven2001revival}. In this paper we argue that a similar phenomenon may be observed in on-going experiments on a charge Kondo device~\cite{iftikhar2017tunable}. 

As introduced in 1980, the two-channel Kondo  (2CK) model~\cite{nozieres1980kondo} describes a single impurity spin $\vec{S}$ coupled to two electronic channels $\alpha=1,2$ via the spin-flip interaction
\begin{equation}
H_{K}=J \sum_{\alpha =1,2} \psi^\dagger_{\alpha \uparrow}(0) \psi_{\alpha \downarrow}(0) S^+ + H.c..
\end{equation}
While a \emph{single} channel of electrons can completely screen the impurity spin, the presence of two (or more) competing channels turns this model into a paradigmatic example of frustration and non-Fermi liquid (NFL) behavior, with possible broader significance in bulk systems such as heavy fermion materials. Due to the spin-flip process $H_K$, the number of electrons from each channel $\alpha=1,2$ and each spin $\sigma = \uparrow , \downarrow$, $N_{\alpha \sigma} = \int dx \psi^\dagger_{\alpha \sigma}(x) \psi_{\alpha \sigma}(x)$, may change but only by unit steps. As a precursor to fractionalization in this model, through the Emery-Kivelson (EK) solution~\cite{emery1992mapping}, one introduces charge, spin, flavor, and spin-flavor quantum numbers,
\begin{eqnarray}\label{csfx}
\mathcal{N}_{{\rm{c,s}}} &=& \frac{1}{2}(N_{1\uparrow} \pm N_{1\downarrow}+N_{2\uparrow} \pm N_{2\downarrow}), \\ \nonumber
\mathcal{N}_{{\rm{f,sf}}} &=& \frac{1}{2}(N_{1\uparrow} \pm N_{1\downarrow}-N_{2\uparrow} \mp N_{2\downarrow}), 
\end{eqnarray}
\begin{figure}[t]
	\centering
	\includegraphics[scale=0.42]{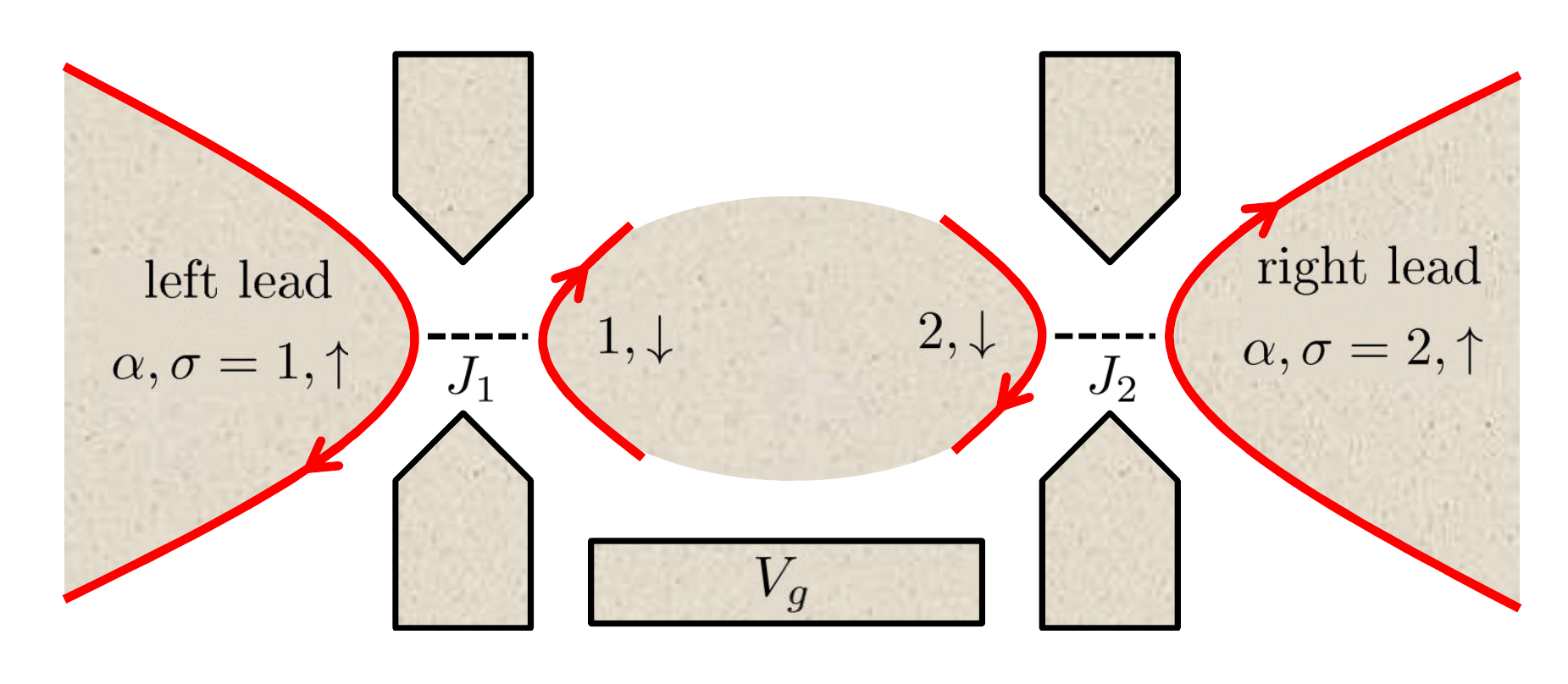}
	\caption{Schematics of the device: two QPCs coupled to a large quantum dot, with couplings constants $J_1,J_2$. The overall charge of the dot is controlled by a gate voltage $V_g$.     }
	\label{fig1}
\end{figure}
and associated new fermions, $\psi^\dagger_\mu(x)$ (here $\mu={\rm{c,s,f,sf}}$), that change only the corresponding $\mathcal{N}_{\mu}$ quantum numbers by a unit step. An exact rewriting of the Kondo interaction is~\cite{emery1992mapping} $H_K = J [(\psi_\mathrm{s}(0) \psi_{\mathrm{sf}}^\dagger(0)) + (\psi_\mathrm{s}(0)\psi_{\mathrm{sf}}(0))] S^+ + H.c.$. Crucially, at weak coupling physical operators such as $H_K$, involve the new fermions \emph{in pairs}~\cite{von1998finite,zarand1998simple}. This is a necessary constraint to describe Fermi liquid (FL) states, since upon inverting Eq.~(\ref{csfx}), each single new fermionic particle changes electronic numbers $N_{\alpha \sigma}$ by \emph{half-integers}~\cite{von1998finite,zarand1998simple}; for example $\psi_{\mathrm{sf}}^\dagger$ takes $\mathcal{N}_{\mathrm{sf}} \to \mathcal{N}_{\mathrm{sf}}  +1$ or equivalently $\delta (N_{1\uparrow},N_{1\downarrow},N_{2\uparrow},N_{2\downarrow}) = (\frac{1}{2},-\frac{1}{2},-\frac{1}{2},\frac{1}{2})$. In this sense, the new fermions are ``confined" to occur in pairs in physical processes in FLs. However, the non-perturbative Kondo interaction leads to non-Fermi liquid behavior~\cite{nozieres1980kondo}. In light of this, one may wonder - \emph{is unpairing of these  fermions possible and can it be manifest in a physical system?}

In recent years the multichannel Kondo effect was experimentally studied in highly tunable semiconductor quantum dot systems ~\cite{potok2006observation,mebrahtu2012observation,keller2015universal,iftikhar2016two,iftikhar2017tunable}. Our work is primarily motivated by charge 2CK setups, theoretically suggested my Furusaki and Matveev~\cite{PhysRevB.52.16676} and recently realized in the quantum Hall regime~\cite{iftikhar2016two}. The impurity ``spin" is encoded by two nearly degenerate macroscopic charge states of a large quantum dot, see Fig.~\ref{fig1}, which is coupled to normal leads via quantum point contacts (QPCs) allowing to flip the ``spin" via single electron tunneling. Upon
decreasing temperature below the Kondo temperature $T_K$ the conductance reaches half of the conductance quantum $G \to \frac{1}{2} \frac{e^2}{h}$, corresponding to two perfectly transmitting quantum resistors in series. 

We study non-equilibrium transport through such devices and analyze the non-linear current $I(V)$ and shot noise $S(V)$, focusing on the vicinity of the 2CK critical point. Generally, shot noise informs on the charge of the current carrying particles, examples ranging from the fractional quantum Hall effect~\cite{de1998direct,PhysRevLett.79.2526} to superconductor junctions exhibiting Cooper pair tunneling~\cite{PhysRevLett.90.067002}. Applying methods borrowed from Gogolin and Komnik \cite{gogolin2006towards} and Schiller and Hershfield \cite{schiller1995exactly}, we find interesting universal properties encoded in the current and noise in the non-equilibrium Kondo regime $ T \ll eV  \ll T_K$ (for simplicity we set $T=0$ for now). The current $I  = \frac{e^2 V}{2 h} ( 1- e|V|/ T_K )+ \mathcal{O} (V^3)$ contains a non-linear correction that corresponds to a backscattering current $I_b = \frac{e^2 V}{2 h} \frac{e|V|}{T_K}$. While the first term describes noiseless current through two perfectly transmitting QPCs, the backscattering current produces shot noise $S =2 e^* I_b + \mathcal{O} (V^3)$, with a Fano-factor
$e^*=e/2$.

This fractional Fano factor can be precisely interpreted in terms of unpairing of a spin-flavor fermion $\psi^\dagger_{\rm{sf}}$ in physical processes at the NFL state: we identify the elementary backscattering processes consisting of annihilation of this individual fermion which yield half-integer changes in electronic occupation numbers, as directly reflected in the shot noise. We clarify how this unpaired fermion appears in the emergent leading irrelevant operator of dimension $3/2$, first identified within the conformal field theory (CFT) exact solution of the Kondo effect~\cite{affleck1995conformal}. Finally we suggest an alternative three lead setup~\cite{iftikhar2017tunable} to probe this fractionalization.

\emph{Model.--} Our system in Fig.~\ref{fig1} consists of a large metallic quantum dot in the quantum Hall regime with spinless electrons coupled to two normal leads via QPCs and described by the Hamiltonian~\cite{PhysRevB.52.16676,mitchell2016universality}
\begin{equation}
\label{H}
\begin{split}
H_{\text{K}}=&\sum_{\alpha=1,2} \Big [i \hbar v_F \sum_{\sigma} \int dx \psi^\dagger_{\alpha \sigma} (x) \partial_x \psi_{\alpha \sigma}(x)\\
& + J_{\alpha} \left ( \psi_{\alpha \uparrow}^{\dagger}(0)\psi_{\alpha \downarrow }^{\phantom{\dagger}}(0) \hat{S}^{-} + H.c. \right )\Big ]  +\Delta E \hat{S}^z .  
\end{split}
\end{equation}
Here, $\sigma=\uparrow$ describes states in the lead and $\sigma=\downarrow$  in the dot; the index $\alpha=1,2$ labels the two QPCs. We assume~\cite{PhysRevB.52.16676,iftikhar2016two} that the dot is sufficiently large, such that its level spacing is small
compared to the temperature, as a result of which edge states in the dot near different QPCs are incoherently coupled. We specialize to the large charging energy limit such that only two macroscopic charge states, with $N = N_0$ or $N_0+1$ electrons in the dot, are relevant  at the experiment's temperatures $T \ll E_c$, and play the role of the impurity spin $S$. Upon detuning the gate voltage from the degeneracy point, an energy splitting $\Delta E$ is formed between these macroscopic charge states. 

The two-channel Kondo state is a critical point occurring at charge degeneracy $\Delta E=0$ and for left-right symmetry $J_1=J_2=J$, which will be assumed. Towards the end we will comment on deviations from these conditions which lead to a crossover at low energies to a FL state~\cite{PhysRevB.52.16676,mitchell2016universality}. The parameters of the model include the density of states $\nu$ and a high-energy cutoff $D$, set by the minimum of the band-width and the charging energy, defining through the tunneling amplitude $J$ the Kondo temperature $T_K \sim D e^{-\frac{1}{\nu J}}$. 

The model Eq.~(\ref{H}) is an anisotropic $XY$ Kondo Hamiltonian with the $J_z$ term omitted. In our calculations below we will add such a term $H_z =J_z \sum_{\alpha , \sigma,\sigma'} \psi^\dagger_{\alpha \sigma }(0)\frac{\vec{\sigma}_{\sigma \sigma'}}{2} \psi_{\alpha \sigma' }(0) S^z $, keeping in mind that spin anisotropy does not affect the low energy physics~\cite{hewson1997kondo}.

\emph{Strategy.--} 
\label{se:shotnoise}
Our primary interest is in the non-equilibrium transport properties and specifically on the shot noise in the vicinity of the 2CK fixed point.
As a non-perturbative tool allowing to approach the vicinity of the strong coupling 2CK fixed point, we use the EK solution near the Toulouse point $J_z = 2 \pi \hbar v_F$. The analysis \emph{at} the Toulouse point gives correctly only the fixed point properties such as the $T=V=0$ value of the linear conductance $G \to G_0=\frac{e^2}{2h}$. However this free fermion description misses the leading low energy corrections. It is well known that for multichannel Kondo models these arise from the leading irrelevant operator known from CFT~\cite{affleck1995conformal}. We clarify that in the 2CK model the leading irrelevant operator is turned on via any small  deviation from the Toulouse point. Thus, treating $J_z - 2 \pi \hbar v_F$ as a perturbation, allows us to perform a controlled \emph{non-equilibrium} calculation in terms of the original electronic degrees of freedom, and capture the leading low energy corrections at the 2CK fixed point.

\emph{Mapping to the Toulouse Hamiltonian.--}
Following the standard EK transformation~\cite{emery1992mapping,schiller1995exactly} we (i) bosonize the fermionic fields $\psi_{\alpha\sigma}(x)\sim \frac{1}{\sqrt{2 \pi a}} e^{i\Phi_{\alpha\sigma}(x)}$, with $a$ being a short distance cutoff, (ii) perform the rotation in Eq.~(\ref{csfx}) to define charge, spin, flavor, and spin-flavor bosons, $\Phi_{\alpha\sigma} \to \Phi_{\mu}$ ($\mu = {\rm{c,s,f,sf}}$), and finally (iii) refermionize these bosons into new fermion operators $\psi_\mu \sim \frac{1}{\sqrt{2 \pi a}} e^{i \Phi_{\mu}}$. The 
transformed Hamiltonian becomes
\begin{eqnarray}
\label{HK2}
 \nonumber
H_K &=& i\hbar v_{F}\sum_{{\rm{\mu}}}\int dx\psi_{\mu}^{\dagger}\left(x\right)\partial_x\psi_{\mu}\left(x\right)+ i \mathcal{J} \chi_{\rm{sf}}(0) \hat{b} \nonumber \\ 
&-&
\frac{eV}{2} \int dx \left[\psi_{{\rm{sf}}}^{\dagger} \left(x\right)\psi_{{\rm{sf}}} \left(x\right)+\psi_{{\rm{f}}}^{\dagger} \left(x\right)\psi_{{\rm{f}}} \left(x\right)\right] \nonumber\\ 
&+&i (J_z-2\pi\hbar v_F) \psi^\dagger_\mathrm{s}(0) \psi_\mathrm{s}(0) \hat{a} \hat{b},
\end{eqnarray}
where $\hat{a},\hat{b}$ are a local Majorana operators associated with the impurity degrees of freedom, $i \hat{a} \hat{b}=S^z$, satisfying $\hat{a}^2=\hat{b}^2=\frac{1}{2}$, $\chi_{\rm{sf}}(x) = \frac{\psi_{\mathrm{sf}}^{\dagger}\left(x\right)+\psi_{\mathrm{sf}}\left(x\right)}{\sqrt{2}}$, and $\mathcal{J}= \frac{J_1+J_2}{\sqrt{2 \pi a}} $. We included the source-drain voltage $eV$, setting a chemical potential difference in the leads $eV \frac{N_{1 \uparrow} - N_{2 \uparrow}}{2}$, which after the transformation Eq.~(\ref{csfx}) simply becomes a chemical potential of the new fermions~\cite{schiller1995exactly}. The last term accounts for deviations from the Toulouse point. The current operator is given by $\hat{I}=\frac{i e}{\hbar}[\frac{N_{1 \uparrow} - N_{2 \uparrow}}{2},H_K]$.

Applying this free fermion Hamiltonian \emph{at the Toulouse point}, as detailed in the appendix, one obtains the current $I(V)=\frac{e^2}{2 h}V(1 + \mathcal{O}(\frac{V^2}{T_K^2}))$ and noise $S(V)= \mathcal{O}(V^3/T_K^2)$, where $ T_K = \pi \nu \mathcal{J}^2$, valid for $eV \ll T_K$. As noted above, the 2CK fixed point $eV / T_K \to 0$ corresponds to two perfectly transmitting QPCs in series which do not produce any partitioning noise~\cite{blanter2000shot} and thus we have $S=0$. The quadratic voltage corrections in $I(V)$ and the cubic term in $S(V)$ are artifacts of the free fermion resonant level structure. 
We shall now obtain the leading universal corrections in $eV/T_K$ that emerge due to deviations from the free fermion point.

\emph{Irrelevant operator and shot noise near the critical point.--} From the CFT solution of the multichannel Kondo effect~\cite{affleck1995conformal} one can identify the leading irrelevant operator, which captures the low energy corrections around the strong coupling fixed point. For the 2CK model this is the dimension 3/2 operator~\cite{affleck1995conformal}, which can be written in terms of EK fermions as~\cite{maldacena1997majorana}
\begin{equation}
\label{pot}
 H_{\mathrm{irr}} = \frac{1}{\nu^{3/2}\sqrt{T_K}}i \psi^\dagger_\mathrm{s}(0) \psi_\mathrm{s}(0) \tilde{\chi}_{\mathrm{sf}}(0)\hat{a}.
\end{equation}
Here, $\nu = \frac{1}{2 \pi \hbar v_F}$ is the density of states, $T_K$ acts as a high energy scale, and $\tilde{\chi}_{\mathrm{sf}}(x) = \chi_{\mathrm{sf}}(x) {\rm{sign}} (x)$ is a modified spin-flavor Majorana fermion, reflecting the absorption of the local Majorana fermion $\hat{b}$ as detailed in the appendix. How does the source-drain voltage couple to $\tilde{\chi}_{\mathrm{sf}}$? 
Answering this question requires one to
formulate an approach fully in terms of the original degrees of freedom in Eq.~(\ref{HK2}), in which the voltage enters in a simple way. 

We obtain such a controlled approach by treating the deviations from the Toulouse point $v_1=J_z-2\pi\hbar v_F$ in Eq.~(\ref{HK2})  perturbatively. To zeroth order in $v_1$, the Majorana operator $\hat{b}$ gets hybridized with the field $ \chi_{\mathrm{sf}}(x)$. Solving this quadratic model one obtains the operator relation~\cite{PhysRevLett.102.047201,PhysRevB.79.125110}
\begin{equation}
\label{eq:bchi}
\hat{b}=\frac{1}{\sqrt{\pi \nu T_K}}\tilde{\chi}_{\mathrm{sf}}(0),
\end{equation} 
which becomes exact at energies $\ll T_K$. Consequently,
\begin{equation}
i v_1 \psi^\dagger_\mathrm{s}(0) \psi_\mathrm{s}(0) \hat{a}  \hat{b}  =-i \frac{v_1}{\sqrt{\pi \nu T_K}} \psi^\dagger_\mathrm{s}(0) \psi_\mathrm{s}(0) \tilde{\chi}_{\mathrm{sf}}(0) \hat{a}.
\label{coupling}
\end{equation}  
Thus, by setting the deviation from the anisotropic Toulouse point to $v_1 = - \sqrt{\pi}/ \nu$, we generate the leading dimension $3/2$ irrelevant operator Eq.~(\ref{pot}) at the 2CK fixed point. As described in detail in the appendix, a  straight forward though tedious computation of the current as well as shot noise in the framework of Eq.~(\ref{HK2}), to infinite order in $\mathcal{J}$ and to the leading second order in $v_1/{\sqrt{\pi \nu T_K}}$, gives our main results
\bea
\label{result}
I &=& \frac{e^2V}{2 h}\left( 1- \frac{\pi^2}{8} \frac{e\left|V\right|}{ T_K} +\mathcal{O}\left(\frac{eV}{T_K}\right)^2\right),  \nonumber \\
S &=&  \frac{e^3V}{2 h} \left( \frac{\pi^2}{8} \frac{e\left|V\right|}{T_K} +\mathcal{O} \left(\frac{eV}{T_K}\right)^2\right).
\eea
Notably, we can define the backscattering current $I_b\equiv G_0V - I = \frac{e^2 V}{h} \frac{\pi^2 e|V|}{16 T_K}$, and write the shot noise as $S = 2 e^* I_b$ with $e^* = e/2$.

\begin{figure}[t]
	\centering
	\includegraphics[scale=0.42]{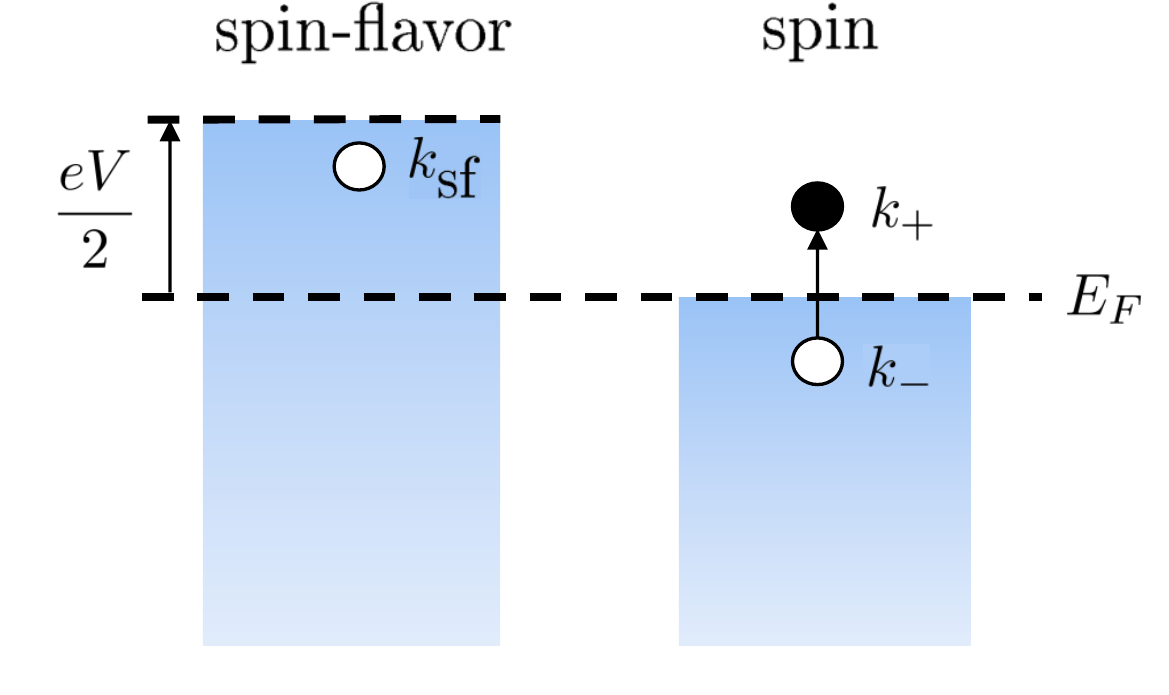}
	\caption{Energy diagram of EK fermions. Elementary backscattering processes consist of annihilation of a single spin-flavor fermion accompanied by a creation of a particle-hole excitation in the spin Fermi sea.}
	\label{fg:3}
\end{figure}

The same result, supplemented by an intelligible physical picture, can be obtained by a simple calculation based on the Fermi's golden rule applied directly with respect to the irrelevant operator Eq.~(\ref{pot}). Decomposing the operator $\tilde{\chi}_{\mathrm{sf}}(0)=\frac{1}{ \sqrt{2L}} \sum_{k_{\mathrm{sf}}}(\tilde{c}^\dagger_{k_{\rm{sf}}}+\tilde{c}_{k_{\rm{sf}}})$ into normal fermionic modes~\cite{zarand1998simple}, we see that it either creates a particle or a hole in the spin-flavor Fermi sea. Eq.~(\ref{HK2}) shows that the source-drain voltage sets an enhanced chemical potential $eV/2$ of the spin-flavor Fermi sea. Thus annihilation of one spin-flavor particle at $k_{\rm{sf}}$ above the equilibrium Fermi level $0 <\varepsilon_{k_{\rm{sf}}} < eV/2$ lowers the energy. Energy conservation is attained via a creation of a particle-hole excitation in the spin sector via the factor $\psi_\mathrm{s}^\dagger(0)\psi_\mathrm{s}(0)= \frac{1}{L} \sum_{k_+}\sum_{k_-}c^\dagger_{\mathrm{s},k_+}c_{\mathrm{s},k_-}$ in Eq.~(\ref{pot}). This process depicted in Fig.~\ref{fg:3} creates a unit change in the quantum numbers in Eq.~(\ref{csfx}), $\tilde{\mathcal{N}}_{\mathrm{sf}} \to \tilde{\mathcal{N}}_{\mathrm{sf}}-1$, \emph{i.e.}, it annihilates an unpaired spin-flavor fermion. Evaluating the total rate for this process  via Fermi's golden rule gives, using $\hat{a}^2=\frac{1}{2}$,
\bea
-\frac{d\left\langle \tilde{\mathcal{N}}_{{\rm{\mathrm{sf}}}}\right\rangle}{dt}  &=&\frac{2\pi}{\hbar L^{3}} \sum_{k_\pm ,k_{\mathrm{sf}}}\frac{1}{\pi \nu T_K} \theta(\epsilon_{k_+})\theta(\epsilon_{k_-})\theta(eV/2-\epsilon_{k_{\mathrm{sf}}})  \nonumber \\
& \times & \delta(\epsilon_{k_{+}}-\epsilon_{k_{-}}-\epsilon_{k_{\mathrm{sf}}})= \frac{\pi}{\hbar}\frac{(eV)^2}{16T_K}.\\ \nonumber
\eea
Crucially, the unit change in $ \tilde{\mathcal{N}}_{\mathrm{sf}}$, modifies \emph{electronic} occupations by half integers. Thus this Poissonian process describes backscattering of charge $e^* = e/2$, and the backcattering current is
\begin{equation}
I_b=-e^*\frac{d \left \langle \tilde{\mathcal{N}}_{\mathrm{sf}}\right\rangle}{dt}= \frac{ e^2V}{h}\frac{\pi^2e|V|}{16T_K},\\ \nonumber
\end{equation}
with an associated noise $S = 2 e^* I_b$, in agreement with Eq.~(\ref{result}). 

\emph{Finite temperature effects.--}  The universality of our results Eq.~(\ref{result}) is revealed by the ratio between the coefficients of  $eV/T_K$  in the current and noise. Another experimentally testable universal ratio can be obtained from the leading $T$ dependence of the current, which we find to be $I  = G_0 V \left[ 1- \frac{\pi^2}{8} \left(\frac{e\left|V\right|}{ T_K} +\pi^2 \frac{T}{T_K} \right) \right]$. Similarly, the noise $S(V,T)$ has a temperature dependence where, at $T \gg eV$ it must cross from the shot noise limit Eq.~(\ref{result}) to thermal noise $S = 4  k_B T G$ with $G=dI/dV|_{V=0}$. 

\emph{Deviations from the critical point.--} We first test the influence of \emph{relevant} perturbations.  The  intricate properties of the critical point are destabilized by left-right asymmetry $\Delta J = J_1 - J_2$ or by gate voltage deviations from the charge degeneracy point $\Delta E$. These create an energy scale, $T^* = c_1 T_K (\nu \Delta J)^2 + c_2 (\Delta E)^2/T_K$, with $c_{1,2}$ coefficients of order unity. Below this energy scale the system crosses over to a FL state, whereby the non-linear conductance gradually decreases below $G_0=e^2/2h$~\cite{PhysRevB.52.16676,mitchell2016universality} till it vanishes at $eV \ll T^*$.  Since $T_K$ may become high and approach the charging energy ($\sim 290mK$) in charge-kondo devices~\cite{iftikhar2016two,iftikhar2017tunable} one may realistically assume $T^* \ll T_K$. This gives a finite voltage window $T^* \ll  eV  \ll T_K$ within which our shot noise predictions, dominated by the leading irrelevant operator, hold. Nevertheless, what is the leading influence of finite $T^*$?   Including for instance channel asymmetry, and assuming $eV \ll T_K$, the current and noise Eq.~(\ref{result}) acquire the corrections~\cite{appendix}
\bea
\label{I_corr}
\delta I &=& - \frac{e}{h} T^* \arctan \frac{eV}{2 T^*}, \nonumber \\
\delta S&=& \frac{e^2}{h} T^* \int_{-eV/2T^*}^{eV/2T^*}   dy\frac{1}{1+y^2} \left( 1-\frac{1}{1+y^2} \right),
\eea
which are valid for any ratio  $T^*/eV$. This current remains a small backscattering correction compared to $G_0 V$ for $eV \gg T^*$. Under this condition Eq.~(\ref{I_corr}) gives $\delta I =- \frac{ e}{ h} \frac{\pi}{2} T^*+ \mathcal{O}(V^{-1})$, and $\delta S = \frac{e^2}{h} \frac{\pi}{2} T^* + \mathcal{O}(V^{-1})$, and the total current and noise are $I+\delta I$ and $S+ \delta S$. Interestingly, to leading order in $T^*$ the fractional Fano factor $e^*/e =(S+ \delta S )/(2e (G_0 V -I-\delta I)) =1/2 $ reappears. This is expected since, similar to the irrelevant operator Eq.~(\ref{pot}), also the dimension $1/2$ relevant channel asymmetry operator consists of one unpaired spin-flavor fermion $(\psi^\dagger_{\rm{sf}} - \psi_{\rm{sf}})/(\sqrt{2}i)$, or another $\psi_\mu$ fermion for other relevant perturbations \emph{e.g.} $\Delta E$~\cite{PhysRevLett.106.147202}.  Thus, although relevant operators eventually destabilize the critical point, the $e^*/e=1/2$ Fano factor is remarkably stable and includes the leading effects of relevant perturbations as well.

We herein validate the stability of our results in presence of generic \emph{marginal} operators. These are quadratic forms of the original electrons $\psi^\dagger_{\alpha \sigma} \psi_{\alpha' \sigma'}$, which map into quadratic forms of the new fermions $\psi^{(\dagger)}_{\mu} \psi^{(\dagger)}_{\nu}$. Such operators, \emph{if present}, lead to corrections to the current due to single electron processes and hence will affect the Fano factor. First, consider $\psi^\dagger_{\alpha \uparrow} \psi_{\alpha' \downarrow}+h.c.$ whereby one electron moves between a lead and the dot. At the free fermion fixed point, this marginal operator changes the charge of the dot, hence, it must involve $S^\pm = ( \hat{a} \mp i\hat{b}) / \sqrt{2}$. Using Eq.~(\ref{eq:bchi}), the latter becomes the unpaired fermion $\tilde{\chi}_{\mathrm{sf}}$, changing electronic numbers in the leads by half integers. Thus, exactly like the marginal operator describing deviations from the Toulouse point, any marginal operator involving the impurity spin $\vec{S}$ at the free fermion fixed point changes into a dimension 3/2 operator at the critical point. Secondly, consider marginal operators of the form $\psi^\dagger_{\alpha \sigma} \psi_{\alpha' \sigma'}$, which, for $\sigma = \sigma'$, do not involve $\vec{S}$. These operators either do not create any charge transfer for $\alpha = \alpha'$, or correspond to an elastic cotunneling processes through both QPCs for $\alpha \ne \alpha'$. Thus, the only process that can yield a charge of $e^*=e$ is a direct tunneling from one lead to the other $\psi^\dagger_{1 \uparrow} \psi_{2 \uparrow} +h.c.$. However, in our large quantum dot with small level spacing such coherent processes are suppressed~\cite{PhysRevB.52.16676}, making our main result of $e^*/e=1/2$ highly stable.

\begin{figure}[b]
	\centering
	\includegraphics[scale=0.48]{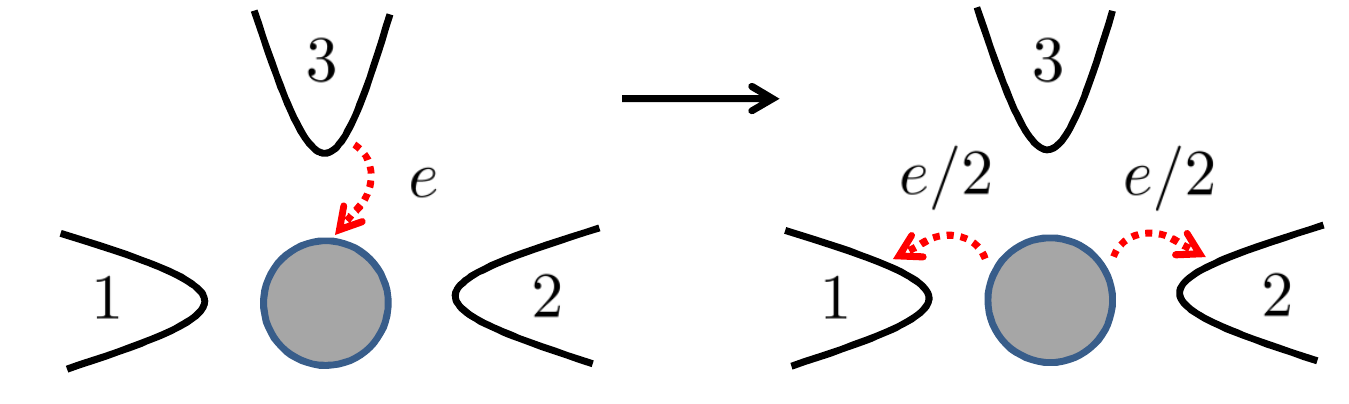}
	\caption{Charge fractionalization using a weak probe: one electron tunnels from the weakly coupled lead no. $3$ and is equally and simultaneously partitioned into the two leads.     }
\end{figure}

\emph{Three-lead setup.--} We briefly present an alternative setup that displays charge fractionalization. Consider attaching a third lead~\cite{iftikhar2017tunable} at voltage $V$ and {\it weakly} coupling it to the large dot with the two other leads held at $V=0$, see Fig.~3. The model Hamiltonian is Eq.~(\ref{H}) where now $\alpha=1,2,3$ and $J_1 = J_2>J_3$. We analyzed the current and noise at the 2CK fixed point for this device~\cite{Inpreparation}.
The injected current through the third QPC is given by $I_3(V) = \frac{e^2 V}{\hbar} \frac{\pi^2}{2} (\nu J_3)^2 [1+ \mathcal{O}(eV/T_K)]$. In the two strongly coupled leads $\alpha,\beta=1,2$, where by channel symmetry the average ejected current is $I \equiv I_\alpha = I_3/2$, we find the current-current correlation $S_{\alpha \beta}(\omega)\simeq 
2 e^* I$, with $e^* = e/2$.

Interpreting this result, each individual charge $e$ tunneling through the 3-rd QPC into the dot, is equally and simultaneously partitioned  into both leads, see Fig.~3. This fractionalization can also be understood along the above Fermi golden rule's picture: the tunneling process from the weakly coupled lead increases the dot's charge, $N \to N+1$, \emph{i.e.}, it involves the operator $S^+ = ( \hat{a} - i\hat{b}) / \sqrt{2}$. Using Eq.~(\ref{eq:bchi}), this operator becomes the unpaired fermion $\tilde{\chi}_{\mathrm{sf}}$, changing electronic numbers in the leads by half integers.

This setup has the advantage that the charge fractionalization can be seen in the shot noise of the full current as measured in one of the leads, rather than in the backscattering current.

\emph{Summary.--}
We analyzed non-equilibrium transport properties of charge 2CK devices and found that shot noise encodes fractional charges signaling key NFL features. While shot noise properties of multichannel Kondo systems have so far remained experimentally elusive, our predictions could now be tested.

One may compare to the single-channel spin-Kondo effect in quantum dots where a Fano factor $e^*/e=5/3$ has been predicted based on a FL theory~\cite{PhysRevLett.97.086601,golub2006shot,PhysRevLett.97.016602,vitushinsky2008effects,mora2008current} and tested~\cite{zarchin2008two,delattre2010noisy,PhysRevLett.106.176601,ferrier2015universality,egger2009kondo}. This is a weighted average of $1e$ and $2e$ processes, while our current result $e^*=e/2$ can not be accommodated within such a Fermi liquid picture. In contrast to charge-Kondo devices~\cite{iftikhar2016two,iftikhar2017tunable}, calculations of non-equilibrium transport in spin-multichannel Kondo devices~\cite{potok2006observation,keller2015universal} remain challenging, but expectedly doable for 2CK devices, due to the free fermion effective description. Also, non-equilibrium noise in $N>2$ multichannel charge Kondo devices~\cite{iftikhar2017tunable} whose linear transport properties were addressed recently~\cite{bao2017quantum} reamins an interesting question for future work. For this theoretical task, connections \emph{e.g.} to topological Kondo devices ~\cite{PhysRevLett.109.156803,altland2013multiterminal,PhysRevLett.113.076401,michaeli2016electron,landau2017two} and their non-equilibrium properties~\cite{zazunov2014transport,PhysRevLett.119.027701} may become useful.

{\it Acknowledgements:} We thank A. Mitchell, C. Mora,  and Y. Oreg for helpful and interesting discussions.


\bibliographystyle{apsrev4-1}

\newpage

\appendix

\section{Full counting statistics (FCS)}
\label{apendix1}
This appendix is devoted to a controllable and detailed derivation of the main results based on the full counting statistics (FCS) method following Gogolin and Komnik \cite{gogolin2006towards}. 
In Sec.~\ref{apendix1a} we start by recapitulating the main definitions of the FCS generating function and its relation to Keldysh Green functions (GF); we apply these definitions in Sec.~\ref{apendix1b} at the Toulouse point as well as including relevant perturbations generating the energy scale $T^*$; finally we apply the FCS method in  Sec.~\ref{apendix1c} perturbatively in the deviations from the Toulouse point providing a controlled derivation of our main result Eq.~(\ref{result}).
\subsection{Preliminaries}
\label{apendix1a}
We define a generating function $\chi(\lambda)=\sum_q e^{iq\lambda}P_q$, where $P_q$ is the probability for transfer of charge $q$ through our system within the measurement time $\mathcal{T}$. Then the cumulants are given by
\begin{equation}
\label{cum}
\left\langle \delta^n q\right\rangle = (-i)^2 \frac{\partial^n}{\partial \lambda^n} \rm{ln}\chi(\lambda)\biggr\rvert_{\lambda=0}.
\end{equation}
The generating function is given by the following average on the Keldysh contour $C$ \cite{levitov2004counting} 
\begin{equation}
\label{}
\chi(\lambda) = \left\langle T_C {\rm{exp}}\left[-\frac{i}{\hbar}\int_C T_\lambda(t) d t\right] \right\rangle,
\end{equation}
where $T_C$ is the contour ordering operator, and $\lambda(t)$ is a contour dependent measuring field which is non-zero only during the measurement time. The operator $T_\lambda(t)$ here describes direct tunneling between the left and right leads. It is coupled to the measuring field $\lambda$ with the generic form
\begin{equation}
\label{}
T_\lambda(t) =e^{i\lambda/2}T_R +e^{-i\lambda/2}T_L, 
\end{equation}
where $T_R$ and $T_L$ are operators transferring an electron through the system to the right or to the left, respectively. In fact the measuring field enters via a gauge transformation
\be
\label{gauge}
\psi_L \to \psi_L e^{i \lambda/4},~~~\psi_R \to \psi_R e^{-i \lambda/4},
\ee
where $\psi_{L/R}$ annihilates a particle in the left$/$right leads.
The generating function can be expressed as ${\rm{ln}} \chi =-i\mathcal{T} \mathcal{U}(\lambda,-\lambda)$, in terms of a field $\mathcal{U}$ which satisfies 
\begin{equation}
\label{fh}
\frac{\partial}{\partial \lambda_-} \mathcal{U}(\lambda_-,\lambda_+) = \left\langle \frac{\partial H(\lambda)}{\partial \lambda_-}\right\rangle_\lambda.
\end{equation}
Here, $-$ and $+$ denote the forward and backward parts of the Keldysh contour.
Thus, on evaluating Eq.~(\ref{fh}), integrating the result over $\lambda_-$ and finally setting $\lambda_-=-\lambda_+=\lambda$, one obtains the generating function for the cumulants. The current and noise are then obtained from the first two cumulants,
\begin{equation}
\label{current}
I= \frac{e \langle \delta q \rangle}{\mathcal{T}} ,~~~S= \frac{2e^2 \langle \delta^2 q \rangle}{\mathcal{T}}.
\end{equation}
For a weak current composed of uncorrelated tunneling events, the effective charge $e^*$ is $e^*/e= \left\langle \delta^2 q\right\rangle / \left\langle \delta q\right\rangle$.

\subsection{FCS at the Toulouse point}
\label{apendix1b}
Our next step is to incorporate the measuring field $\lambda(t)$ into the Toulouse Hamiltonian (\ref{HK2}). This is done by generalizing the gauge transformation Eq.~(\ref{gauge}) to include the dot-states, $\sigma = \downarrow$, which remain gauge invariant,  
\begin{eqnarray}
\psi_{1,\uparrow} &\rightarrow& e^{i\lambda(t)/4}\psi_{1,\uparrow} \\ \nonumber
\psi_{2,\uparrow} &\rightarrow& e^{-i\lambda(t)/4}\psi_{2,\uparrow} \\ \nonumber
\psi_{\alpha,\downarrow} &\rightarrow& \psi_{\alpha,\downarrow},~~~(\alpha=1,2). \\ \nonumber
\end{eqnarray}
Going through the EK transformation, one can show that this gauge transformation maps to 
\begin{eqnarray}
\label{gt}
\psi_{\mathrm{c,s}} \rightarrow  \psi_{\mathrm{c,s}},  ~~~
\psi_{\mathrm{f,sf}} \rightarrow  e^{i\lambda(t)/4}\psi_{\mathrm{f,sf}}.  \nonumber
\end{eqnarray}
Thus, the tunneling term $\mathcal{J}$ in Eq.~(\ref{HK2}) transforms to 
\begin{equation}
H_\lambda = i\mathcal{J} \hat{b}\left[\chi_{\mathrm{sf}}\mathrm{cos}(\lambda(t)/4)+\eta_{\mathrm{sf}}\mathrm{sin}(\lambda(t)/4)\right],
\end{equation}
where $\chi_{\mathrm{sf}}=\frac{\psi_{\mathrm{sf}}^{\dagger}+\psi_{\mathrm{sf}}}{\sqrt{2}}$ and $\eta_{\mathrm{sf}}=\frac{\psi_{\mathrm{sf}}^{\dagger}-\psi_{\mathrm{sf}}}{i\sqrt{2}}$ are the Majorana fermions associated with the spin-flavor field.
We observe that the gauge transformation Eq.~(\ref{gt}) can be identified with a rotation of the Majorana components. Defining a rotated Majorana basis
\begin{equation}
\label{trans}
\left(
\begin{array}{c}
\chi_{\mathrm{sf}}^\lambda\\
\eta_{\mathrm{sf}}^\lambda\\
\end{array}
\right) =
\left(
\begin{array}{cc}
\mathrm{cos}(\lambda(t)/4) & \mathrm{sin}(\lambda(t)/4) \\
-\mathrm{sin}(\lambda(t)/4)& \mathrm{cos}(\lambda(t)/4)  \\
\end{array}
\right) 
\left(
\begin{array}{c}
\chi_{\mathrm{sf}}\\
\eta_{\mathrm{sf}}\\
\end{array}
\right),
\end{equation} 
the Toulouse Hamiltonian takes the form
\begin{eqnarray}
H_\lambda &=& i\hbar v_{F}\sum_{\mu=\mathrm{c},\mathrm{s},\mathrm{f}}\int dx  \psi^\dagger_{\mu}\left(x\right) \partial_x \psi_{\mu}\left(x\right) \\ \nonumber
&+& \frac{i}{2}\hbar v_{F}\int\mathrm{d}x\left[\chi_{\mathrm{\mathrm{sf}}}^\lambda\left(x\right)\partial_x\chi_{\mathrm{sf}}^\lambda\left(x\right)+\eta_{\mathrm{sf}}^\lambda\left(x\right)\partial_x\eta_{\mathrm{sf}}^\lambda\left(x\right)\right] \\ \nonumber
&+& i\mathcal{J}\chi_{\mathrm{sf}}^\lambda\hat{b}+i\mathcal{J}_-\eta_{\mathrm{sf}}^\lambda\hat{a},
\label{Hlambda}
\end{eqnarray}
where $\mathcal{J}_-= \frac{J_1-J_2}{\sqrt{2 \pi a}} $. We first address the symmetric lead couplings setup, where $\mathcal{J}_-=0$. In this case, note that only $\chi_{\mathrm{sf}}^\lambda$ is coupled to the impurity, and the RHS of Eq.~(\ref{fh}) is
\begin{equation}
\left\langle \frac{\partial H_{\lambda}}{\partial\lambda^{-}}\right\rangle =\frac{i\mathcal{J}}{4}\left\langle \hat{b}\left[\eta_{\mathrm{sf}}\mathrm{cos}(\lambda(t)/4)-\chi_{\mathrm{sf}}\mathrm{sin}(\lambda(t)/4)\right]\right\rangle.
\end{equation}
Expressing the above equation in terms of Keldysh GF's (for the form of the electronic free GF's, see Eq.~(34) in the paper of Gogolin and Komnik \cite{gogolin2006towards} and replace $V \to V/2)$), we obtain
\begin{equation}
\label{ceq}
\begin{split}
\left\langle \frac{\partial H_{\lambda}}{\partial\lambda^{-}}\right\rangle =\frac{i\Gamma}{4} \int \frac{\mathrm{d}\omega}{2\pi} \bigg\{  D_{bb}^{--}(\omega)(n_2-n_1)  \\  
  + D_{bb}^{-+}(\omega)\left[e^{i\bar{\lambda}/4}(1-n_2)-e^{-i\bar{\lambda}/4}(1-n_1)\right] \bigg\} , 
\end{split}
\end{equation}
where we define $\Gamma=T_K= \pi \nu \mathcal{J}^2$, $\bar{\lambda}=\lambda^--\lambda^+$, $D_{bb}(t)=-i\left\langle T_C \hat{b}(t)\hat{b}(0)\right\rangle$ is the full GF of the $\hat{b}$ operator and $n_{1,2}(\omega)$ are Fermi-Dirac functions of leads $\alpha=1,2$ respectively. Taking the advantage of the quadratic structure of this Hamiltonian, one can exactly calculate~\cite{gogolin2006towards}
\begin{eqnarray}
\label{Dbb}
&&D_{bb}(\omega) =\frac{i\Gamma}{\mathrm{Det}(g_0^{-1}-\Sigma_K)}\times\\ \nonumber 
&&\left[
\begin{array}{cc}
\frac{i\omega}{\Gamma}+n_1+n_2-1 & e^{i\bar{\lambda}/4}n_1+e^{-i\bar{\lambda}/4}n_2 \\
-e^{i\bar{\lambda}/4}(1-n_2)-e^{-i\bar{\lambda}/4}(1-n_1) & -\frac{i\omega}{\Gamma}+n_1+n_2-1  \\
\end{array}
\right],
\end{eqnarray} 
where
\begin{equation}
\begin{split}
-\mathrm{Det}(g_0^{-1}-\Sigma_K)=\omega^2+\Gamma^2+\Gamma^2\left[n_1(1-n_2)(e^{i\bar{\lambda}/2}-1)\right. \\  
  \nonumber
 \left.  +n_2(1-n_1)(e^{-i\bar{\lambda}/2}-1)\right].
\end{split}
\end{equation}
Plugging this in Eq. (\ref{ceq}) yields the following generating function
\begin{equation}
\label{lnchi}
\begin{split}
\mathrm{ln} \chi = \frac{\mathcal{T}}{2\hbar} \int \frac{\mathrm{d}\omega}{2\pi}\mathrm{ln} \bigg\{  1+T(\omega)\left[n_1(1-n_2)(e^{i\lambda}-1)\right. \\  
 \left.  + n_2(1-n_1)(e^{-i\lambda}-1)\right] \bigg\} \;, 
\end{split}
\end{equation}
where $T(\omega)=\frac{\Gamma^2}{\Gamma^2+\omega^2}$. 

Focusing on low energies $\omega \ll T_K$, we see that $T(\omega) \to 1$. The $\lambda$-dependence $e^{\pm i \lambda}$ signifies single electron transport processes. At $T=0$ the generating function becomes $\mathrm{ln} \chi =\mathcal{T}\frac{i\lambda eV}{2h}$, such that only the first moment is finite. Consequently, we obtain conductance $G=\frac{e^2}{2h}$, together with noise $S=0$. This happens naturally in the limit $eV,k_B T \ll T_K$ where the system is exactly at the fixed point.

Before looking at the irrelevant operator, we would now like to include the effect of channel asymmetry $\mathcal{J}_-= \frac{J_1-J_2}{\sqrt{2 \pi a}} $ and observe the resulting  current and noise. The change $\delta \left\langle \frac{\partial H_{\lambda}}{\partial\lambda^{-}}\right\rangle$ in Eq.~(\ref{Hlambda}) is  
\begin{equation}
\label{ert}
\delta \left\langle \frac{\partial H_{\lambda}}{\partial\lambda^{-}}\right\rangle =-\frac{i\mathcal{J}_-}{4}\left\langle \hat{a}\left[\chi_{\mathrm{sf}}\mathrm{sin}(\lambda(t)/4)+\eta_{\mathrm{sf}}\mathrm{cos}(\lambda(t)/4)\right]\right\rangle.
\end{equation}  
Due to the similar structure of the equation one can write Eq.~(\ref{ceq}) for the asymmetry term by replacing $\Gamma\rightarrow \Gamma_-$, $\bar{\lambda}\rightarrow -\bar{\lambda}$, $D_{bb}(\omega)\rightarrow D_{aa}(\omega)$, where $\Gamma_-=\pi \nu \mathcal{J}_-^2$, and $D_{aa}(t)=-i\left\langle T_C \hat{a}(t)\hat{a}(0)\right\rangle$ is the full GF of the $\hat{a} $ operator. The generating function then has the exact same form as Eq.~(\ref{lnchi}), only with $\lambda\rightarrow -\lambda$ and $T(\omega)=\frac{\Gamma_-^2}{\omega^2+\Gamma_-^2}$.  It is then easy to obtain Eq.~(\ref{I_corr}) at $T=0$. Note that the change in the sign of $\lambda$ in the form $e^{-i \lambda}$  indicates a negative contribution to the current.


\subsection{Deviations from the Toulouse point}
\label{apendix1c}
We now consider the effect of irrelevant operators Eq.~(\ref{pot}) emerging at the vicinity of the 2CK fixed point on the generating function and its cumulants. We have clarified that the irrelevant operator is generated by the deviations from the Toulouse point, see Eq.~(\ref{coupling}).
Thus, by evaluating Eq. (\ref{fh}) in the presence of $v_1$ as a perturbation we expect to obtain at low energies the same behavior of the dimension $3/2$ irrelevant operator.

In order to calculate the generating function in the presence of the coupling Eq.~(\ref{coupling}) one has to evaluate perturbative corrections to the $\hat{b}$-impurity's GF . To lowest order in $v_1$, the impurity's GF is 
\begin{equation}
\bar{D}_{bb}= D_{bb} + v_1^2 D_{bb}\Sigma D_{bb} = D_{bb}+\delta D_{bb},
\end{equation}
where $\Sigma$ is the self energy. Thus, Eq. (\ref{ceq}) acquires the correction
\begin{equation}
\label{ceq2}
\begin{split}
\left\langle \frac{\partial \delta\mathcal{U}}{\partial\lambda^{-}}\right\rangle =\frac{i\Gamma}{4} \int \frac{\mathrm{d}\omega}{2\pi} \bigg\{ \delta D_{bb}^{--}(\omega)(n_2-n_1) \\  
 + \delta D_{bb}^{-+}(\omega)\left[e^{i\bar{\lambda}/4}(1-n_2)-e^{-i\bar{\lambda}/4}(1-n_1)\vphantom{\int}\right] \bigg\}. 
\end{split}
\end{equation}
The self-energy takes the form
\begin{equation}
\Sigma^{ij}(\omega) = \int \frac{\mathrm{d}\omega_1}{2\pi} \hat{d}_{aa}^{ij}(\omega-\omega_1) \int  \frac{\mathrm{d}\omega_2}{2\pi} G^{ji}_\mathrm{s}(\omega_1+\omega_2)G_\mathrm{s}^{ij}(\omega_2),
\end{equation}
where $\hat{d}_{aa}=-i\left\langle T_C \hat{a}(t)\hat{a}(0)\right\rangle$ and $G_\mathrm{s}=-i\left\langle T_C\psi^\dagger_\mathrm{s}(t)\psi_\mathrm{s}(t)\right\rangle$ are bare $\hat{a}$-GF's of the impurity and the spin fermionic operators, respectively. At $T=0$ these two different GF's take the form 
\begin{equation}
G_s^{ij}(\omega) =
2\pi\nu\left[
\begin{array}{cc}
-\frac{i}{2} \mathrm{sign}(\omega) &  i\Theta (-\omega) \\
-i\Theta (\omega) &  -\frac{i}{2} \mathrm{sign}(\omega) \\
\end{array}
\right],
\end{equation} \begin{equation}
\hat{d}_{aa}^{ij}(\omega) =
\left[
\begin{array}{cc}
1/\omega & i\pi \delta(\omega) \\
-i\pi \delta(\omega) &  -1/\omega \\
\end{array}
\right].
\end{equation} 
Evaluating $\Sigma^{ij}(\omega)$ at $T=0$, we obtain
\begin{equation}
\Sigma^{ij}(\omega) =
\nu^2\omega\left[
\begin{array}{cc}
\mathrm{ln}\left|\omega\right|-1 & -i\pi \Theta (-\omega) \\
-i\pi \Theta (\omega) &  -\mathrm{ln}\left|\omega\right|+1 \\
\end{array}
\right].  
\end{equation} 
To obtain $D_{bb}(\omega)$ at low energies, we take the limit $\omega \ll \Gamma $ of Eq.~(\ref{Dbb}),
\begin{eqnarray}
\label{Dbb2}
&D_{bb}&(\omega) \to \frac{i\Gamma}{\mathrm{Det}(g_0^{-1}-\Sigma_K)_{\omega\rightarrow 0}}\\ \nonumber 
&\times&\left[
\begin{array}{cc}
n_1+n_2-1 & e^{i\bar{\lambda}/4}n_1+e^{-i\bar{\lambda}/4}n_2 \\
-e^{i\bar{\lambda}/4}(1-n_2)-e^{-i\bar{\lambda}/4}(1-n_1) & n_1+n_2-1  \\
\end{array}
\right].
\end{eqnarray} 
We find that the integral over $\delta D_{bb}^{--}$ in  Eq. (\ref{ceq2}) vanishes. Looking at the second term  $\delta D_{bb}^{-+}=\sum_{i,j}D_{bb}^{-i}\Sigma^{ij} D_{bb}^{j+}$ in detail, we find
\begin{eqnarray}
&D_{bb}^{--}&\Sigma^{--} D_{bb}^{-+} =- D_{bb}^{-+}\Sigma^{++} D_{bb}^{++},\\ \nonumber
&D_{bb}^{--}&\Sigma^{-+} D_{bb}^{++}  = 0, 
\end{eqnarray}
such that the only contribution to the generating function comes from the term
\begin{equation}
D_{bb}^{-+}\Sigma^{+-} D_{bb}^{-+}=i\pi\nu^2\omega\Theta (\omega)\left[\frac{\Gamma(e^{i\bar{\lambda}/4}n_1+e^{-i\bar{\lambda}/4}n_2)}{\mathrm{Det}(g_0^{-1}-\Sigma_K)_{\omega\rightarrow 0}}\right]^2.
\end{equation}
Plugging this term into Eq. (\ref{ceq2}), we obtain the correction for the generating function at $T=0$  
\begin{equation}
\label{gf}
\mathrm{ln} \delta\chi = \frac{\mathcal{T} }{\hbar\Gamma} \left(\frac{v_1\nu eV}{4}\right)^2e^{-i\lambda/2}. 
\end{equation}
Note the factor $1/2$, to be reflected in the fractional Fano factor, and the negative sign of $\lambda$ which indicates a negative correction to the current,
\begin{equation}
I = \frac{e^2V}{2h}\left[1- \pi\nu^2 v_1^2\frac{e\left|V\right|}{8\Gamma}\right].  
\end{equation}
Defining the backscattering current $I_b \equiv I-\frac{e^2V}{4\pi \hbar}$ we are easily able to calculate the cumulants of Eq.~(\ref{gf}) and the value of the effective charge 
\begin{equation}
e^*/e= \left(\frac{\partial^2 \mathrm{ln} \delta\chi }{\partial \lambda^2}\biggr/\frac{\partial \mathrm{ln} \delta\chi }{\partial \lambda}\right)\biggr\rvert_{\lambda=0}=\frac{1}{2}.
\end{equation}

\section{Derivation of Eq.~(\ref{eq:bchi})}
\label{apendix2}
 This short appendix provides a derivation of Eq.~(\ref{eq:bchi}) following Refs.~\onlinecite{PhysRevLett.102.047201,PhysRevB.79.125110}.  We examine the exact behavior of the coupled operators $\chi_{{\rm{sf}}}(x)$ and $\hat{b}$ in the vicinity of the fixed point as described by Eq.~(\ref{HK2}) in the absence of the perturbation Eq.~(\ref{pot}). Consider the mode expansion
\begin{eqnarray}
\chi_{\mathrm{sf}}(x) &=& \sum_k \varphi_k(x) \psi_k + \rm{H.c} \\ \nonumber 
\hat{b} &=& \sum_k u_k \psi_k +\rm{H.c},
\end{eqnarray}
where $\psi_k$ are operators in Fock-space satisfying $\left\{\psi_k,\psi^\dagger_{k'}\right\}=\delta_{k,k'}$, $ \varphi_k(x)$ are wave functions and $u_k$
are local coefficients. In this basis, the Hamiltonian (\ref{HK2}) reads $H'=\sum_k \varepsilon_k \psi^\dagger_k \psi_k$ such that the wave functions $\varphi_k(x)$ and $u_k$ satisfy a set of Schrodinger equations $\left[H_K,\chi_{\mathrm{sf}}(x)\right]=\left[H',\chi_{\mathrm{sf}}(x)\right]$, $\left[H_K,\hat{b}\right]=\left[H',\hat{b}\right]$ yielding 
\begin{eqnarray}
&i \mathcal{J} \delta(x)u_k+i\hbar v_F \partial_x\varphi_k(x) = \varepsilon_k\varphi_k(x), \\ \nonumber 
&-i \mathcal{J} \varphi_k(0) = \varepsilon_k u_k.
\end{eqnarray}
Solving these equations, one obtains $\varphi_k(x)\propto e^{ikx}[\theta(x)\varphi_k^{(+)}+\theta(-x)\varphi_k^{(-)}]$, $\varphi_k(0)=\frac{1}{2}(\varphi_k^{(+)}+\varphi_k^{(-)})$, $u_k=\frac{i \mathcal{J}}{\hbar v_F k}\varphi_k(0)$, $\varphi_k^{(+)}/\varphi_k^{(-)}=e^{-2i\mathrm{tan}^{-1}\left[\frac{1}{\hbar v_F k}\left(\frac{\mathcal{J}^2}{\hbar v_F  }\right)\right]}$. Thus, for energies $\ll T_K$ one finds $\varphi_k^{(+)}=-\varphi_k^{(-)}$. Although the wave functions   has a discontinuity at $x=0$, the local Majorana $\hat{b}$ field can be written as \cite{PhysRevLett.102.047201,PhysRevB.79.125110}
\begin{equation}
\hat{b}=\frac{1}{\sqrt{\pi \nu T_K}}\tilde{\chi}_{\mathrm{sf}}(0),
\end{equation}  
where $\tilde{\chi}(x)=\chi_{\mathrm{sf}}(x)\mathrm{sign}(x)$ is continuous at $x=0$. 

\end{document}